\documentclass[a4paper,12pt]{article}
\usepackage{amsmath}
\usepackage{pstricks}
\usepackage{pst-node}
\usepackage[ansinew]{inputenc}
\usepackage{amssymb,amsmath}
\usepackage{amsfonts}
\usepackage{epsfig}

\def\p{\partial}

\def\a{\alpha}
\def\b{\beta}
\def\g{\gamma}
\def\d{\delta}
\def\o{\omega}

\font\Sets=msbm10

    \def\Real {\hbox{\Sets R}}

   \def\Natural {\hbox{\Sets N}}

\def\be{\begin{equation}}       \def\ba{\begin{array}}

\def\ee{\end{equation}}         \def\ea{\end{array}}

\def\bea {\begin{eqnarray}}      \def\eea {\end{eqnarray}}

\def\bean{\begin{eqnarray*}}    \def\eean{\end{eqnarray*}}

\def\la  {\lambda}

\def\eps{\varepsilon}           

\def\const {\mathop{\rm const}\nolimits}

\def\ker  {\mathop{\rm Ker} \nolimits}

\def\RA {\ \Rightarrow\ }         \def\LRA {\ \Leftrightarrow\ }

\def\qed   {\vrule height0.6em width0.3em depth0pt}

\def\<{\langle} \def\({\left(}  \def\>{\rangle} \def\){\right)}

\newtheorem{exi}{Example}

\author{E. Kartashova, A. Shabat}
\title{Computable Integrability.\\
Chapter 2: Riccati equation}

\begin{document}
\date{}
\maketitle
\tableofcontents {
\newpage
\section{Introduction}
Riccati equation (RE)

\begin{equation}\label{ric}
\boxed{\phi_x=a(x)\phi^2+b(x) \phi+c(x)}
\end{equation}

is one of the most simple nonlinear differential equations because
it is of {\bf first order} and with {\bf quadratic nonlinearity}.
Obviously, this was the reason that as soon as Newton invented
differential equations, RE was the first one to be investigated
extensively since the end of the 17th century \cite{Ric}. In 1726
Riccati considered the first order ODE
             $$w_x=w^2+u(x)$$
with polynomial in $x$ function $u(x).$ Evidently, the cases $\deg
u=1,\, 2$ correspond to the Airy and Hermite transcendent
functions, respectively. Below we show that Hermite transcendent
is integrable in quadratures.  As to Airy transcendent, it is only
F-integrable\footnote{See Ex.3}  though the corresponding equation
itself is at the first glance a simpler one.\\

Thus,  new transcendents  were introduced as solutions of the
first order ODE with the quadratic nonlinearity, i.e. as solutions
of REs. Some classes of REs are known to have general solutions,
for instance:
$$
y^{'} + ay^2=bx^{\alpha}
$$
where all $a, b, \alpha$ are constant in respect to $x$. D.
Bernoulli discovered(1724-25) that this RE is integrable in
elementary functions if $\alpha=-2$ or $\alpha=-4k(2k-1),
k=1,2,3,....$. Below some general results about RE are presented
which make it widely usable for numerous applications in different
branches of physics and mathematics.

\section{General solution of RE}
In order to show how to solve (\ref{ric}) in general form, let us
regard two cases.
\subsection{$a(x)=0$}

In case $a(x)=0$, RE takes particular form

\be \label{ric1}\phi_x=b(x) \phi+c(x), \ee

i.e. it is a first-order LODE and its general solution can be
expressed in quadratures. As a first step, one has to find a
solution $z(x)$ of its homogeneous part\footnote{see Ex.1}, i.e.
$$
z(x): \quad z_x=b(x) z.
$$
In order to find general solution of Eq.(\ref{ric1}) let us
introduce new variable $\tilde{\phi}(x)=\phi(x)/z(x)$, i.e.
$z(x)\tilde{\phi}(x)=\phi(x)$. Then
$$
(z(x)\tilde{\phi}(x))_x =b(x)z(x)\tilde{\phi}(x)+ c(x), \quad
\mbox{i.e.} \quad z(x)\tilde{\phi}(x)_x=c(x),
$$
and it gives us general solution of Eq.(\ref{ric1}) in quadratures

\be \label{varcon}\phi(x)=z(x)\tilde{\phi}(x)=z(x)(\int
\frac{c(x)}{z(x)}dx +\const). \ee

This method is called {\bf method of variation of constants} and
can be easily generalized for a system of first-order LODEs
$$
\vec{y}^{'}=A(x)\vec{y}+\vec{f}(x).
$$
Naturally, for the system of $n$ equations we need to know $n$
particular solutions of the corresponding homogeneous system in
order to use method  of variation of constants. And this is
exactly the bottle-neck of the procedure -
 in distinction with first-order LODEs which are all
integrable in quadratures, already second-order LODEs {\bf are not}.\\

\subsection{$a(x) \neq 0$}
In this case {\bf one known particular solution of a RE allows
to construct its general solution}.\\

Indeed, suppose that $\varphi_1$ is a particular solution of
Eq.(\ref{ric}), then
$$
c= \varphi_{1,x}-a\varphi_1^2-b\varphi_1
$$
and  substitution $\phi=y+\varphi_1$ annihilates free term $c$
yielding to an equation
 \be \label{help1} y_x=ay^2+\tilde{b}y \ee
with $\tilde{b}=b+2a\varphi_1.$ After re-writing Eq.(\ref{help1})
as
$$
\frac{y_x}{y^2}=a+ \frac{\tilde{b}}{y}
$$
and making an obvious change of variables $\phi_1=1/y$, we get a
particular case of RE
$$
\phi_{1,x}+{\tilde{b}}\phi_{1}-a=0
$$
and its general solution is written out explicitly in the previous
subsection.\\

\paragraph{Example 2.1 }

As an important illustrative example leading to many applications
in mathematical physics, let us regard a particular RE in a form
\be \label{Hermit} y_x + y^2= x^2 + \a. \ee

For $\a=1$,  particular solution can be taken as $y=x$ and general
solution obtained as above yields to
$$
y=x+\frac{e^{-x^2}}{\int e^{-x^2} d x + \const},
$$
i.e. in case (\ref{Hermit})  is integrable in quadratures.
Indefinite integral $\int e^{-x^2} d x$ though not expressed in
elementary functions, plays important role in many areas from
probability
theory till quantum mechanics.\\

For arbitrary $\a$,   Eq.(\ref{Hermit}) possess remarkable
property, namely, after an elementary fraction-rational
transformation \be \label{Dirac}\hat{y}=x+\frac{\a}{y+x} \ee it
takes form
$$
\hat{y}_x+\hat{y}^2=x^2+\hat{\a}, \ \  \hat{\a}=\a+2,
$$
i.e. form of original Eq.(\ref{Hermit}) did not change while its
rhs increased by 2. In particular, after this transformation
Eq.(\ref{Hermit}) with $\a=1$ takes form

$$
\hat{y}_x+\hat{y}^2=x^2+3
$$

and since $y=x$ is a particular solution of (\ref{Hermit}), then
$\hat{y}=x+1/x$ is a particular solution of the last equation. It
means that for any $$ \a=2k+1, \ \ k=0,1,2,...$$ general solution
of
Eq.(\ref{Hermit}) can be found in quadratures as it was done for the case $\a=1$. \\

In fact, it means that Eq.(\ref{Hermit}) is {\bf form-invariant}
under the transformations  (\ref{Dirac}). Further we are going to
show that general RE possess similar property as well.

\subsection{Transformation group}

Let us check that general fraction-rational change of variables

\begin{equation}\label{group}
\hat{\phi}=\frac{\a (x) \phi + \b (x)}{\g (x)\phi + \delta (x)}
\end{equation}

 transforms one Riccati equation into the another one similar to
Example 2.1. Notice that (\ref{group}) constitutes group of
transformations generated by

$$
\frac{1}{\phi} ,\quad  \a (x) \phi, \quad \phi +\b (x),
$$
thus only actions of generators have to be checked:\\

\begin{itemize}
\item{} $\hat{\phi}=1/\phi$ transforms (\ref{ric})  into \\
$$\hat{\phi}_x+c(x)\hat{\phi}^2+b(x)\hat{\phi}+a(x)=0,$$
\item{} $\hat{\phi}=\a (x) \phi$ transforms (\ref{ric})  into \\
$$\hat{\phi}_x-\frac{a(x)}{\a (x)}\hat{\phi}^2- [b(x)+(\log
\a(x))_x]\hat{\phi}-\a (x) c(x)=0,$$
\item{} $\hat{\phi}=\phi +\b (x)$ transforms (\ref{ric})  into\\
$$\hat{\phi}_x-a(x)\hat{\phi}^2+[2\b (x) a(x)- b(x)]\hat{\phi} - \hat{c}= 0,$$
where $$\hat{c}=a(x)\b^2(x)- b(x)\b(x)+c(x)+\b(x)_x.$$
\end{itemize}
Thus, having {\bf one solution} of a some Riccati equation we can
get immediately general solutions of the whole family of REs
obtained from the original one under the action of transformation
group
(\ref{group}).\\

It is interesting to notice that for Riccati equation knowing {\bf
any three solutions} $\phi_1,\phi_2,\phi_3$ we can construct all
other solutions $\phi$ using a very simple formula called {\bf
cross-ratio}:
\be
\label{cross}\frac{\phi-\phi_1}{\phi-\phi_2}=A\frac{\phi_3-\phi_1}{\phi_3-\phi_2}
\ee
with an arbitrary constant $A$, where choice of $A$ defines a
solution. In order to verify this formula let us notice that
system of equations

$$
\begin{cases}
\dot{\phi}=a(x)\phi^2+b(x)\phi+c(x)\\
\dot{\phi}_1=a(x)\phi_1^2+b(x)\phi_1+c(x)\\
\dot{\phi}_2=a(x)\phi_2^2+b(x)\phi_2+c(x)\\
\dot{\phi}_3=a(x)\phi_3^2+b(x)\phi_3+c(x)
\end{cases}
$$
is consistent if
$$
\begin{bmatrix}
\dot{\phi}&\phi^2&\phi&1\cr \dot{\phi}_1&\phi_1^2&\phi_1&1\cr
\dot{\phi}_2&\phi_2^2&\phi_2&1\cr
\dot{\phi}_3&\phi_3^2&\phi_3&1\cr
\end{bmatrix}=0
$$
and direct calculation shows that this condition is equivalent to

\be \label{Dratio}
\frac{d}{dx}\big(
\frac{\phi-\phi_1}{\phi-\phi_2} \cdot
\frac{\phi_3-\phi_1}{\phi_3-\phi_2}\big)=0.
\ee

 As it was shown, REs  are
not invariant under the action of (\ref{group}) while
(\ref{group}) conserves the form of equations but not form of the
coefficients. On the other hand, it is possible to construct  new
differential equations related to a given  RE which will be
invariant with respect to transformation group (\ref{group}) (see
next section).\\

At the end of this section we consider a very interesting example
\cite{Adler} showing  connection of Eq.(\ref{Dratio}) with first
integrals for
generalization of one of Kovalevskii  problems \cite{Kov}.\\

\paragraph{ Adler´s example} System of equations
\be \label{AdlerKov} y_{j,x}+2y_j^2=sy_j, \quad s=\sum_{j=1}^n
y_j, \quad j=1,2,...n
 \ee
was studied by Kovalevskii in case $\boxed{n=3}$ and it was shown
that there exist two quadratic first integrals
$$
F_1=(y_1-y_2)y_3, \quad F_2=(y_2-y_3)y_1
$$
and therefore  Kovalevskii problem  is integrable in
quadratures.\\

In case $\boxed{n \ge 4}$ the use of (\ref{Dratio}) gives us
immediately following some first integrals
$$
\frac{y_l-y_i}{y_l-y_j} \frac{y_k-y_i}{y_k-y_j},
$$
i.e. Sys.(\ref{AdlerKov}) has nontrivial first integrals for
arbitrary $n$.\\

 It is interesting that for this example solution of
Sys.(\ref{AdlerKov}) is easier to construct without using its
first integrals. Indeed, each equation of this system is a Riccati
equation if $a$ is regarded as given, substitution
$y_i=\phi_{i,x}/2\phi_{i}$ gives
$$\phi_{i,xx}=s\phi_{i,xx}, \ \phi_{i,xx}=a(x)+c_i, \
s=a_{xx}/a_x$$ and equation for $a$ has form
$$
\frac{a_{xx}}{a_x}=\frac{a_{x}}{2}(\frac{1}{a-c_1}+...+\frac{1}{a-c_n}).
$$
After integration $a_x^2=\const(a-c_1)...(a-c_n)$, i.e. problem is
integrable in quadratures (more precisely, in hyper-elliptic
functions).\\

In fact, one more generalization of Kovalevskii problem can be
treated along the same lines - case when function $s$ {\bf is not}
sum of $y_j$ but some arbitrary function $s=s(x_1,...,x_n)$. Then
equation on $a$ takes form
$$
\frac{a_{xx}}{a_x}=a_{x}s(\frac{a_{x}}{a-c_1}+...+\frac{a_{x}}{a-c_n})
$$
which concludes Adler´s example.

\subsection{Singularities of solutions}

All the properties of Riccati equations which have been studied
till now, are in the frame of local theory of differential
equations.  We just ignored possible existence of singularities of
solutions regarding all its properties locally, in a neighborhood
of a point. On the other hand, in order to study analytical
 properties of solutions, one needs to know character of
 singularities, behavior of solutions at infinity,
 etc.\\

 One can distinguish between two main types of singularities -
 singularities,
 not depending on initial conditions (they are called {\bf fixed}) and
depending on initial conditions (they are called {\bf movable}).
Simplest possible singularity is a pole, and that was the reason
why first attempt of classification of the ordinary nonlinear
differential equations of the first and second order, suggested by
Painleve, used this type of singularities as criterium. Namely,
list of all equations was written out, having only {\bf poles as
movable singularities} (see example of P1 in Chapter 1), and nice
analytic properties of their solutions have been found. It turned
out that, in particular, Painleve equations  describe self-similar
solutions of solitonic equations (i.e. equations in partial
derivatives): P2 corresponds to KdV (Korteweg-de Vries equation),
P4 corresponds to NLS (nonlinear Schrödinger equation)
and so on. \\

Using cross-ratio formula (\ref{cross}), it is easy to demonstrate
for a Riccati equation that {\bf all  singularities} of the
solution $\phi$, with an exception  of singularities of particular
solutions $\phi_1,\phi_2,\phi_3$, {\bf are movable poles}
described as following:
$$
\phi_3=\frac{1}{1-A}(\phi_2-A\phi_1)
$$
where $A$ is a parameter defining the solution $\phi$. Let us
construct a solution with poles for Eq.(\ref{Hermit}) from Example
2.1. We take a solution in a form \be
\label{series}y=\frac{1}{x+\varepsilon}
+a_0+a_1(x+\varepsilon)+a_2(x+\varepsilon)^2+... \, \ee with
indefinite coefficients $a_i$, substitute it into (\ref{Hermit})
and make equal terms corresponding to the same power of
$(x+\varepsilon)$. The final system of equations takes form
$$
\begin{cases}
a_0  =  0, \\
3a_1-\a - \varepsilon^2=0, \\
4a_2+2\varepsilon=0, \\
5a_3-1+a_1^2=0, \\
6a_4+2a_1a_2=0, \\
7a_5+2a_1a_3+a_2^2=0 \\
 ...
\end{cases}
$$
and in particular for $\a=3, \ \varepsilon=0$ the coefficients are
 $$a_1=1, \ a_2=a_3=...=0$$ which corresponds to the solution
 $$
y=x+\frac{1}{x}
 $$
which was found already in Example 2.1.\\

 This way we have also
learned that each pole of solutions have order 1. In general case,
it is possible to prove that series (\ref{series}) converges  for
arbitrary pair $(\varepsilon, \ \a)$ using the connection of RE
with the theory of linear equations (see next section). In
particular for fixed complex $\a$, it means that for any point
$x_0=-\varepsilon$ {\bf there exist the only solution of
(\ref{Hermit})} with a pole
in this point.\\

As to nonlinear first order differential equations (with
non-quadratic nonlinearity), they have more complicated
singularities. For instance, in a simple example
$$
y_x=y^3+1
$$
if looking for a solution of the form $y=ax^k+....$ one gets
immediately
$$
akx^{k-1}+...=a^3x^{3k}+... \ \ \RA k-1=3k  \ \ \RA 2k=-1
$$
which implies that singularity here is a branch point, not a pole
(also see \cite{Reid}). It make RE also very important while
studying degenerations of Painleve transcendents. For instance,
(\ref{Hermit}) describes particular solutions of P4 (for more
details see Appendix).

\section{Differential equations related to RE}

\subsection{Linear equations of second order}

One of the most spectacular properties of RE is that its theory is
in fact {\bf equivalent} to the theory of second order homogeneous
LODEs
\be \label{hom}
 \psi_{xx}=b(x)\psi_x+c(x)\psi
\ee because it can easily be shown that these equations can be
transformed into Riccati form and {\it viceversa}. Of course, this
statement is only valid if
Eq.(\ref{ric}) has non-zero coefficient $a(x)$, $a(x) \neq 0$.\\

$\blacktriangleright$ Indeed, let us regard second-order
homogenous LODE (\ref{hom}) and make change of variables

$$
\phi=\frac{\psi_x}{\psi}, \quad \mbox{then}
 \quad \phi_x=\frac{\psi_{xx}}{\psi} -
 \frac{\psi_{x}^2}{\psi^2},$$

which implies
$$
 \frac{\psi_{xx}}{\psi}=\phi_x +  \frac{\psi_{x}^2}{\psi^2}  =\phi_x +  \phi^2
$$

and after substituting the results above into initial LODE,  it
takes  form

\begin{equation}
\phi_x=\phi^2+b(x) \phi+c(x).\nonumber
\end{equation}

which is particular case of RE. \qed \\

$\blacktriangleleft$ On the other hand, let us regard general RE
\begin{equation}
\phi_x=a(x)\phi^2+b(x) \phi+c(x)\nonumber
\end{equation}
and suppose that $a(x)$ is not $\equiv 0$ while condition of
$a(x)\equiv 0$ transforms RE into first order linear ODE which can
be solved in quadratures analogously to Thomas equation (see
Chapter 1). Now, following change of variables

$$
\phi = -\frac{\psi_x}{a(x)\psi}
$$

transforms RE into

$$
 -\frac{\psi_{xx}}{a(x)\psi}+
 \frac{1}{a(x)}\Big(\frac{\psi_{x}}{\psi}\Big)^2+
 \frac{a(x)_x}{a(x)^2}\frac{\psi_{x}}{\psi}=
 a(x)\Big(\frac{\psi_{x}}{a(x)\psi}\Big)^2
- \frac{b(x)}{a(x)}\frac{\psi_{x}}{\psi}+
 c(x)
$$

and it  can finally be reduced to

$$
a(x)\psi_{xx}-\big[ a(x)_x + a(x)b(x)\big]\psi_x+ c(x)a(x)^2\psi=0
$$
which is second order homogeneous LODE. \qed\\

Now, analog of the result of Section 2.2 for second order
equations can be proved.

\paragraph{Proposition 3.1}  {\it Using one solution of a second order
homogeneous
LODE, we can construct  general solution as well.}\\

$\blacktriangleright$ First of all, let us prove that without loss
of generality we can put $b(x)=0$ in $
\psi_{xx}+b(x)\psi_x+c(x)\psi=0$. Indeed, change of variables
$$
\psi(x)=e^{-\frac{1}{2}\int b(x)dx} \hat{\psi}(x) \quad
\Rightarrow
\quad\psi_x=(\hat{\psi}_x-\frac{1}{2}b\hat{\psi})e^{-\frac{1}{2}\int
b(x)dx}
$$
and finally \be \label{canon2}
\hat{\psi}_{xx}+\hat{c}\hat{\psi}=0, \quad
\hat{c}=c-\frac{1}{4}b^2-\frac12 b_x.
 \ee

Now, if we know one particular solution $\hat\psi_1$ of
Eq.(\ref{canon2}), then it follows from {the considerations above
that RE
$$
\phi_x+\phi^2+\hat{c}(x)=0
$$
has a solution $\phi_1=\hat{\psi}_{1,x}/\hat{\psi}_1.$ The change
of variables $\hat{\phi} = \phi-\phi_1$ annihilates the
coefficient $\hat{c}(x)$:
$$
(\hat{\phi}+\phi_1)^{'}+\big( \hat{\phi}+\phi_1\big)^2)+
\hat{c}(x)=0  \quad \Rightarrow \quad \hat{\phi}_x+\hat{\phi}^2
+2\phi_1 \hat{\phi}=0\quad \Rightarrow
$$
\be \label{temp} (\frac{1}{\hat{\phi}})_x = 1
+2\phi_1\frac{1}{\hat{\phi}}, \ee

i.e. we reduced our RE to the particular case Eq.(\ref{ric1})
which is integrable in quadratures. Particular solution
$z=1/\hat{\phi}$ of homogeneous part of Eq.(\ref{temp}) can be
found from
$$
z_x=2z\frac{\hat{\psi}_{1,x}}{\hat{\psi}_1} \quad \mbox{as} \quad
z=\hat{\psi}_1^2
$$
and Eq.(\ref{varcon}) yields to

\be \label{psi2} \hat{\psi}_2(x)=\hat{\psi}_1\int
\frac{dx}{\hat{\psi}_1^2(x)}. \ee

Obviously, two solutions $\hat{\psi}_1$ and $\hat{\psi}_2$ are
linearly independent since Wronskian $<\hat{\psi}_1,
\hat{\psi}_2>$ is non-vanishing\footnote{see Ex.2}:
$$<\hat{\psi}_1, \hat{\psi}_2>:= \left| \ba{cc}\hat{\psi}_1 \
\hat{\psi}_2\\\hat{\psi}_1^{'} \ \hat{\psi}_2^{'}\ea \right|=
\hat{\psi}_1 \hat{\psi}_2^{'}-\hat{\psi}_2 \hat{\psi}_1^{'}= 1\ne
0.
$$

Thus their linear combination gives general solution of  Eq.(\ref{hom}).\qed\\

\paragraph{Proposition 3.2} Wronskian $<\psi_1, \psi_2>$ is
constant {\bf iff} $\psi_1$ and $\psi_2$ are solutions of \be
\psi_{xx}=c(x)\psi .\ee

$\blacktriangleright$ Indeed, if $\psi_1$ and $\psi_2$ are
solutions, then
$$
(\psi_1 \psi_2^{'}-\psi_2 \psi_1^{'})^{'}=\psi_1
\psi_2^{''}-\psi_2 \psi_1^{''}=c(x)(\psi_1 \psi_2-\psi_1 \psi_2)=0
\ \ \RA $$$$\RA \ \ \psi_1 \psi_2^{'}-\psi_2 \psi_1^{'}=\const.
 $$ \qed

$\blacktriangleleft$ if Wronskian of two functions $\psi_1$ and
$\psi_2$ is a constant,
$$
\psi_1 \psi_2^{'}-\psi_2 \psi_1^{'}=\const \ \ \RA \ \ \psi_1
\psi_2^{''}-\psi_2 \psi_1^{''}=0$$$$ \RA \ \
\frac{\psi_2^{''}}{\psi_2}=\frac{\psi_1{''}}{\psi_1}.
$$
\qed

{\bf Conservation of the Wronskian} is one of the most important
characteristics of second order differential equations and will be
used further for construction of modified Schwarzian equation.

To illustrate  procedure described in Proposition 3.1, let us take
{\bf Hermite equation} \be \label{H2}\o_{xx}-2x\o_x+2 \lambda
\o=0. \ee Change of variables $z=\o_x/\o$ yields to
$$
z_x=\frac{\o_{xx}}{\o}-z^2, \ \ z_x+z^2-2xz+2\lambda=0
$$
and with $y=z-x$ we get finally
$$
y_x+y^2=x^2-2\lambda-1,
$$
i.e. we got the equation studied in Example 2.1 with
$\a=-2\lambda-1$. It means that all solutions of Hermite equation
with positive integer $\lambda$,  $ \ \lambda=n, \ \ n \in
\Natural$ can easily be found while for negative integer $\lambda$
one needs change of variables inverse to (\ref{Dirac}):
$$
y=-x + \frac{\g}{\hat{y}-x}, \ \ (\hat{y}-x)(y+x)=\g, \ \
\g=\hat{\a}-1, \ \ \hat{\a}=\a+2.
$$
It gives us {\bf Hermite polynomials}
$$
\begin{cases}
\lambda=0, \ \ y=-x, \ \ \o=1 \\
\lambda=1, \ \ y=-x+\frac{1}{x}, \ \ \o=2x \\
\lambda=2, \ \ y=-x+\frac{4x}{2x^2-1)} \ \ \o=4x^2-2 \\
..........\\
\lambda=n, \ \ y=-x+\frac{\o_x}{\o}, \ \ \o= H_n(x)=(-1)^n
e^{x^2}\frac{d^n}{dx^n}(e^{-x^2}) \\
........
\end{cases}
$$
as solutions.\\

Notice that the same change of variables

\be \label{log} \phi=\frac{{\psi}_{x}}{\psi} \ee

which linearized original RE, was also used for linearization of
Thomas equation and Burgers equation in Chapter 1. This change of
variables is called {\bf log-derivative} of function $\psi$ or
$D_x \log(\psi)$ and plays important role in many different
aspects of integrability theory, for instance, when solving
factorization
problem. \\

 \paragraph{Theorem 3.3} Linear ordinary differential operator $L$ of order $n$ could be
 factorized with factor of first order , i.e. $L=M\circ(\p_x - a)$
  for some operator $M$, {\bf iff}
  \be \label{ker}
  a = \frac{\psi_x }{\psi}, \quad \mbox{where} \quad  \psi \in \ker(L).
 \ee \\

$\blacktriangleright$  $L=M\circ(\p_x - a), \ a = \psi_x/ \psi$
implies $(\p_x - a)\psi=0$, i.e. $\psi \in \ker(L)$. \qed \\

$\blacktriangleleft$
 Suppose that $\psi_1=1$ is an element of
the $\ker(L)$, i.e. $\psi_1 \in \ker(L)$. It leads to $a=0$ and
operator $L$ has zero free term  and is therefore divisible by
$\p_x$.\\

If constant function $\psi_1=1 $ does not belong to the kernel of
initial operator, following change of variables
$$
\hat{\psi}= \frac{\psi}{\psi_1}
$$
lead us to a new operator \be \label{sim} \hat{L}=f^{-1}L \circ f
\ee which has a constant as a particular solution $\hat{\psi_1}$
for $f=\psi_1$.
\qed \\

\paragraph{Remark.} Operators $L$ and $\hat{L}$ given by (\ref{sim}), are
called {\bf equivalent  operators} and their properties will be
studied
in detailed in the next Chapter.\\

 Notice that Theorem 3.3 is analogous to the Bezout´s theorem on
divisibility criterium of a polynomial: A polynomial $P(z)=0$ is
divisible on the linear factor, $P(z)=P_1(z)(z-a)$,  iff $a$ is a
root of a given polynomial, i.e. $P(a)=0$. Thus, in fact this
classical theorem constructs one to one correspondence between
factorizability and solvability of $L(\psi)=0$. \\

The factorization of differential operators is in itself a very
interesting problem which we are going to discuss in details in
 Chapter 3. Here we will only regard one very simple
example - LODO with constant coefficients
$$
L(\psi):=\frac{d^n \psi}{dx^n}+a_1\frac{d^{n-1}
\psi}{dx^{n-1}}+\ldots+a_n\psi =0.
$$
In this case each root $\lambda_i$ of a {\bf characteristic
polynomial}
$$
\lambda^n+a_1\lambda^{n-1}+\ldots+a_n=0
$$
generates a corresponding first order factor with
$$
\lambda_i= \frac{\psi_x }{\psi}
$$
ant it yields to
$$
\psi_x =\lambda_i \psi \quad \Rightarrow \quad
\psi=c_ie^{\lambda_i x}
$$
and finally
$$
L=\frac{d^n }{dx^n}+a_1\frac{d^{n-1}
}{dx^{n-1}}+\ldots+a_n=(\frac{d}{dx}-\lambda_1)\cdots(\frac{d}{dx}-\lambda_n).
$$
This formula allows us to construct general solution for
$L(\psi)=0$, i.e. for $\psi \in \ker(L)$, of the form
$$
\psi=\sum c_i e^{\lambda_i x}
$$
in the case of all distinct roots of characteristic polynomial.\\

In case of  double roots $\lambda_k$ with  multiplicity $m_k$ it
can be shown that \be \label{multiple} \psi=\sum P_k(x)
e^{\lambda_k x} \ee where degree of a polynomial $P_k(x)$ depends
on the multiplicity of a root, $deg P_k(x) \leq m_k-1$ (cf. Ex.4)

\subsection{Schwarzian equation}

Let us regard again second-order LODE

\begin{equation}\label{second}
\psi_{xx}+b(x)\psi_x+c(x)\psi=0
\end{equation}

and suppose we have two solutions $\psi_1, \psi_2$ of
(\ref{second}). Let us introduce new function $\varphi=
\psi_1/\psi_2$, then

$$
 \varphi_x =
\frac{\psi_{1x}\psi_2-\psi_{2x}\psi_1}{\psi_2^2}, \quad
\varphi_{xx} =
b(x)\frac{\psi_{1x}\psi_2-\psi_{2x}\psi_1}{\psi_2^2}+
2\frac{\psi_{1x}\psi_2-\psi_{2x}\psi_1}{\psi_2^3}\psi_{2x},
$$
which yields

$$
\frac{\varphi_{xx}}{\varphi_x}= - b(x) - 2
\frac{\psi_{2x}}{\psi_2}
$$

and substituting $\phi =\frac{\psi_{2x}}{\psi_2}= (\log \psi_2)_x$
into (\ref{ric}) related to  (\ref{second}) we get finally

\begin{equation}\label{schwarz}
\frac 34 \left(\frac{\varphi_{xx}}{\varphi_x}\right)^2 - \frac 12
\frac{\varphi_{xxx}}{\varphi_x} = c(x).
\end{equation}

Left hand of (\ref{schwarz}) is called {\bf Schwarz derivative} or
just {\bf Schwarzian} and is invariant in respect to
transformation group (\ref{group}) with constant coefficients $\a,
\b, \g, \d$:
$$
\hat{\varphi}= \frac{\a\varphi+\b}{\g\varphi+\d}.
$$
It is sufficient to check only two cases:
$$
\hat{\varphi}=\frac{1}{\varphi} \ \mbox{and} \ \hat{\varphi}=\a
\varphi + \b.
$$
which can be done directly.\\

This equation plays major role in the theory of conform
transformations of polygons \cite{Gol}.

\subsection{Modified Schwarzian equation}

Notice that substitution $\varphi= \psi_1 /\psi_2$ allowed us to
get invariant form of the initial Eq.(\ref{second}). Another
substitution, namely, $\varphi= \psi_1 \psi_2$, leads to similar
equation which differs from classical Schwarzian equation
(\ref{schwarz}) only by a constant term and this is the reason why
we call it {\bf modified Schwarzian equation}. This form of
Schwarzian equation turns out to be useful for a construction of
approximate solutions of Riccati equations with parameter (see
next section). In order to construct modified Schwarzian equation,
we need following Lemma.

\paragraph {Lemma 3.4} Let $\psi_1, \ \psi_2$ are two linear
independent solutions of \be\label{can2} \psi_{xx}=c(x)\psi. \ee
 Then functions
$$
\psi_1^2,\ \ \psi_2^2, \ \ \psi_1 \psi_2
$$
constitute a basis in the solution space of the following third
order equation: \be
\label{***}\varphi_{xxx}=4c(x)\varphi_{x}+2c_x(x)\varphi. \ee

$\blacktriangleright$ Using notations
$$
\varphi_1=\psi_1^2, \ \varphi_2=\psi_2^2, \ \varphi_1=\psi_1
\psi_2,
$$
we can compute Wronskian $\mathcal{W}$ of these three functions
$$\mathcal{W}=<\varphi_1,\varphi_2,\varphi_3>=(\psi_1 \psi_{2,x}- \psi_2
\psi_{1,x})^3=< \psi_1,\psi_2>^3 $$ and use Proposition 3.2 to
demonstrate that $$\mathcal{W}=\const\neq 0,$$ i.e. functions
$\varphi_i$ are linearly independent.

After introducing notations $$\mathcal{V}= < \psi_1,\psi_2> \ \
\mbox{and} \ \ f_j=\frac{\psi_{j,x}}{\psi_j}$$ it is easy to
obtain
$$
 \frac{\mathcal{V}}{\varphi_3}=f_2-f_1,\quad \frac{\varphi_{3,x}}{\varphi_3}=f_2+f_1 $$
which yields to \be \label{ff} f_1=
\frac{\varphi_{3,x}-\mathcal{V}}{2\varphi_3}, \quad f_2=
\frac{\varphi_{3,x}+\mathcal{V}}{2\varphi_3}. \ee

Substitution of these $f_j$ into
$$
f_{j,x}+f_j^2=c(x)
$$
gives \be\label{**}
  4c(x)\varphi^2+\varphi_x^2-2\varphi \varphi_{xx}= \mathcal{V}^2.
\ee
 with $\varphi=\varphi_3$ and differentiation of Eq.(\ref{**})
with respect to $x$ gives Eq.(\ref{***}) and it easy to see that
{\bf equations (\ref{**}) and (\ref{***}) are equivalent}.

Analogous reasoning shows that $\varphi_1, \ \varphi_2$  are also
solutions of Eq.(\ref{***}). \qed \\

Equation (\ref{***}) as well as its equivalent form (\ref{**})
will be used further for construction of approximate solutions of
REs, they also define solitonic hierarchies for KdV and NLS. It
will be more convenient to use (\ref{**}) in  slightly different
form.\\

Let us rewrite (\ref{**}) as \be
  4c(x)+\frac{\varphi_x^2}{\varphi^2}-\frac{2 \varphi_{xx}}{\varphi}=
\frac{\mathcal{V}^2}{\varphi^2} \nonumber
 \ee
and introduce notation $a=1/\varphi$, then \be \label{modSchwarz}
  c(x)=\frac 34 \frac{a_x^2}{a^2}-\frac 12 \frac{a_{xx}}{a}+
\mathcal{V}^2 a^2
 \ee
and compare this equation with Schwarzian equation \be  c(x)=\frac
34 \left(\frac{\varphi_{xx}}{\varphi_x}\right)^2 - \frac 12
\frac{\varphi_{xxx}}{\varphi_x}   \nonumber \ee

one can see immediately why Eq.(\ref{modSchwarz}) is called {\bf
modified Schwarzian equation}.\\

Notice that after the substitution $a=e^{2b}$,  rhs of modified
Schwarzian equation, i.e. {\bf modified Schwarzian derivative,
Dmod}, takes a very simple form

\be Dmod(a):=\frac 34 \frac{a_x^2}{a^2}-\frac 12 \frac{a_{xx}}{a}=
b_{xx}+b_x^2 \nonumber
 \ee
which is in a sense similar to $D_x \log$. Indeed, for
$\psi=e^\varphi$,
$$D_x\log(\psi)=\frac{\psi_x}{\psi}=\varphi_x=e^{-\varphi}\frac{d}{dx}e^{\varphi},$$
while
$$Dmod(e^{2\varphi})=e^{-\varphi}\frac{d^2}{dx^2}e^{\varphi}.$$\\

At the end of this section let us stress the following  basic
fact: we have shown that from some very logical point of view,
first order  nonlinear Riccati equation, second order linear
equation and third order nonlinear Schwarzian equation {\bf are
equivalent}! It gives us freedom to choose the form of equation
which is most adequate for specific problem to be solved.

\section{Asymptotic solutions}

In our previous sections we have studied Riccati equation and its
modifications as classical ordinary differential equations, with
one independent variable. But many important applications of
second order differential equations consist some additional
parameter $\la$, for instance one of the most significant
equations of one-dimensional quantum mechanics takes one of two
forms \bea
\psi_{xx} & = & (\lambda+u)\psi \label{sch} \\
\psi_{xx} & = & (\lambda^2+u_1\lambda+u_2)\psi \label{zs} \eea
where Eq.(\ref{sch}) is called Schrödinger equation and
Eq.(\ref{zs}) can be considered as  modified Dirac equation in
quantum mechanics while in applications to solitonic hierarchies
it is called Zakharov-Shabat equation. Notice that Schrödinger
equation with $u(x)=x^2$ equivalent to Eq.(\ref{H2}). Coefficients
of these two equations have special names  -  $u, u_1, u_2$ are
called {\bf potentials} due to many physical applications and
$\la$ is called {\bf spectral parameter} because of following
reason. Schrödinger equation, being rewritten as
$$
L(\psi) = \lambda \psi, \quad L(\psi)=\psi_{xx} -u \psi
$$
becomes obviously an equation for eigenfunctions of operator $L$
(with appropriate boundary conditions, of course). This operator
is called {\it Schrödinger operator}.\\

For our convenience we name the whole coefficient before $\psi$ as
{\bf generalized potential} allowing it sometimes to be a
polynomial in $\la$ of {\bf any finite degree}. Coming back to
Eq.(\ref{can2}),
the generalized potential is just the function $c(x)$.\\

Now, with the equation having a parameter, problem of its
integrability became, of course, more complicated and different
approaches can be used to solve it. If we are interested in a
solution for all possible values of a parameter $\la$, asymptotic
solution presented by a formal series can always be obtained
(section 4.1) while for some specific exact solutions can be
constructed (sections 4.2, 4.3) in a case of truncated series. It
becomes possible while existence of a  parameter gives us one more
degree of freedom to play with. Cf. with Example 4.2 where exact
solution has been obtained also as a series and its convergence
resulted from the main theorem of the theory of differential
equations on solvability of Cauchy problem. On the other hand,
this solution is valid only for some restricted set of parameter´
values, namely for integer odd $\a$.

\subsection{RE with a parameter $\la$}

Let us  show first that the RE with a parameter $\la$
corresponding to Eq.(\ref{zs}), namely \be \label{zsR}
f_x+f^2=\la^2+u_1\la+u_2, \quad \mbox{with} \quad f=D_x\log(\psi),
\ee has a solution being represented as a formal series.\\

\paragraph{Lemma 4.1} Eq.(\ref{zsR}) has a solution
\be \label{fSeries}f=\la + f_0 +\frac{f_1}{\la}+... \ee where
coefficients $f_j$ are differential polynomials in $u_1$ and
$u_2$.\\

$\blacktriangleright$ After direct  substituting the series
(\ref{fSeries}) into the equation for $f$ and making equal
corresponding coefficients in front of the same powers of $\la$,
we get

$$
\begin{cases}
2f_0=u_1 \\
2f_1+f_{0,x}+f_{0}^2=u_2\\
2f_2+f_{1,x}+2f_{0}f_{1}=0\\
2f_3+f_{2,x}+2f_{0}f_{2}+f_{1}^2=0\\
.....
\end{cases}
$$
and therefore, coefficients of (\ref{fSeries}) are differential
polynomials of potentials $u_1$ and $u_2$. \qed\\

Notice that taking a series \be \label{fSeries-}g=-\la + g_0
+\frac{g_1}{\la}+\frac{g_2}{\la^2}+\frac{g_3}{\la^3} \quad ... \ee
as a form of solution , we will get a different system of
equations on its coefficients $g_i$:\\
$$
\begin{cases}
-2g_0=u_1 \\
-2g_1+g_{0,x}+g_{0}^2=u_2\\
-2g_2+g_{1,x}+2g_{0}g_{1}=0\\
-2g_3+g_{2,x}+2g_{0}g_{2}+g_{1}^2=0\\
 .....
\end{cases}
$$
Solution of Eq.(\ref{zsR}) constructed in Lemma 4.1 yields to the
solution of original Zakharov-Shabat equation (\ref{zs}) of the
form \be \label{psi1W}\psi_1 (x,\la)=e^{\int f(x,\la) dx}=e^{\la
x}(\eta_0(x)+\frac{\eta_1(x)}{\la}+
\frac{\eta_2(x)}{\la^2}+\frac{\eta_3(x)}{\la^3}+....) \ee
 and
analogously, the second solution is \be \label{psi2W}\psi_2
(x,\la)=e^{\int g(x,\la) dx}=e^{-\la
x}(\xi_0(x)+\frac{\xi_1(x)}{\la}+
\frac{\xi_2(x)}{\la^2}+\frac{\xi_3(x)}{\la^3}+....) \ee In fact,
it can be proven that Wronskian $<\psi_1,\psi_2>$ is a power
series on $\lambda$ (see Ex.8) with constant coefficients. Notice
that existence of these two solutions is not enough to construct
general solution of initial Eq.(\ref{zs}) because linear
combination of these formal series is not defined, also
convergence problem has to be considered. On the other hand,
existence of Wronskian in a convenient form allows us to construct
family of potentials giving convergent series for (\ref{psi1W})
and (\ref{psi2W}). We
 demonstrate it at the more simple example,
namely Schrödinger equation (\ref{sch}).\\

Let us regard Schrödinger equation (\ref{sch}), its RE has form
\be \label{schR} f_x+f^2=\la+u, \quad \mbox{with} \quad
f=D_x\log(\psi), \ee and it can be regarded as particular case of
(\ref{zs}), i.e. the series for $f$ yields to \be
\label{fSeriesk}f=k + f_0 +\frac{f_1}{k}+... , \quad \la=k^2,\ee
and $g(x,k)=f(x,-k)$. We see that in case of (\ref{sch}) there
exists a simple way to calculate function $g$ knowing function $f$
and it allows us to construct two  solutions of Schrödinger
equation (\ref{sch}): \be \label{psi1} \psi_1 (x,k)=e^{\int f(x,k)
dx}=e^{k x}(1+\frac{\zeta_1(x)}{k}+
\frac{\zeta_2(x)}{k^2}+\frac{\zeta_3(x)}{k^3}+....)\ee and
$$\psi_2 (x,k)= \psi_1 (x,-k).$$ Substitution of say $\psi_1$ into
(\ref{sch}) gives a recurrent relation between coefficients
$\zeta_i$:

\be \label{recur} \zeta_{j+1,x}=\frac 12 (u\zeta_j-\zeta_{j,xx}),
\quad \zeta_0=1. \ee

In particular,

\be \label{recur1}  u= 2 \zeta_{1,x}\ee

which means that in order to compute potential $u$  it is enough
to know only {\bf one coefficient $\zeta_{1}$} of the formal
series! Below we demonstrate how this recurrent relation helps us
to define potentials corresponding to a given solution.

\paragraph{Example 4.2} Let us regard truncated series corresponding
to the solutions of (\ref{sch})
$$\psi_1=e^{kx}(1+\frac{\zeta_1}{k}), \ \
\psi_2=e^{-kx}(1-\frac{\zeta_1}{k}),$$ then due to (\ref{recur})
$$
u=2\zeta_{1,x}, \ \ \zeta_{1,xx}=2\zeta_{1,x}\zeta_{1}
$$
and Wronskian $\mathcal{W}$ of these two functions has form \be
\label{wron1} \mathcal{W}=<\psi_1, \psi_2> =
-2k+\frac{1}{k}(\zeta_1^2-\zeta_{1,x}). \ee Notice that
$$(\zeta_1^2-\zeta_{1,x})_x=0$$ and it means that $\mathcal{W}$
does not depend on $x$, $\mathcal{W}=\mathcal{W}(k)$.
 Introducing notation $k_1$ for a zero of the Wronskian,
$\mathcal{W}(k_1)=0$, it is easy to see that
$$\zeta_1^2-\zeta_{1,x}=k_1^2$$ which implies that $\psi_1$ and
$\psi_2$ {are solutions} of (\ref{sch}) with
$$
\zeta_1 = k_1- \frac{2k_1}{1+e^{-2k_1(x-x_0)}} =
-k_1\tanh{k_1(x-x_0)}
$$
and potential \be \label{1soliton}
u=-2\frac{(2k_1)^2}{(e^{k_1(x-x_0)}+e^{-k_1(x-x_0)})^2}=
\frac{-2k_1^2}{\cosh^2 (k_1(x-x_0))}, \ee where $x_0$ is a
constant of integration. \\

It is important to understand here that general solution of
Schrödinger equation (\ref{sch}) can be now found as a linear
combination of $\psi_1$ and $\psi_2$ for all  values of a
parameter $\la=k^2$ {\bf with exception} of two  special cases:
$k=0$  and $k=k_1$ which implies
functions $\psi_1$ and $\psi_2$ are {\bf linearly dependent} in these points.  \\


Fig.1 (...)\\

 At the Fig. 1  graph of potential $u$ is shown and it is
easy to see that magnitude of the potential in the point of
extremum is defined by zeros of the Wronskian $\mathcal{W}$. At
the end of this Chapter it will be shown that this potential
represents a solitonic solution of stationary KdV equation, i.e.
{\bf solution of a Riccati equation generates solitons!}

\subsection{Soliton-like potentials}

It this section we regard only Schrödinger equation (\ref{sch})
and demonstrate that generalization of the Example 4.2 allows us
to describe a very important special class of potentials having
solutions in a form of truncated series.\\

\paragraph{Definition 4.3 } Smooth real-valued function  $u(x)$ such that
$$ \ u(x) \to  0 \quad \mbox{for} \quad x \to \pm \infty,$$
is called {\bf transparent potential} if there exist solutions of
Schrödinger equation (\ref{sch}) in a form of truncated series with potential $u(x)$.\\

Another name for a transparent potential is {\bf soliton-like} or
{\bf solitonic} potential due to many reasons. The simplest of
them is just its form  which is a bell-like one and "wave" of this
form was called a soliton by \cite{zabu} and this notion became
one of the most important in the modern nonlinear physics, in
particular while many nonlinear equations have solitonic
solutions.\\

Notice that truncated series $\psi_1$ and $\psi_2$ can be regarded
as polynomials in $k$ of some degree $N$ multiplied by some
exponent (in Example 4.2 we had $N=1$). In particular, it means
that Wronskian $\mathcal{W}=<\psi_1, \psi_2>$ is odd function,
$\mathcal{W}(-k)=-\mathcal{W}(k)$, vanishing at $k=0$ and also it
is a {\bf polynomial} in $k$  of degree $2N+1$: \be \label{wronn}
\mathcal{W}(k)=-2k\prod_1^N (k^2-k_j^2). \ee

As in Example 4.2, functions $\psi_1$ and $\psi_2$ are linearly dependent at the points
$k_j$, i.e.
$$\psi_1(x,k_j)=A_j\psi_2(x,k_j), \quad j=1,2,...,N$$ with some
constant proportionality coefficients  $A_j$.\\

\paragraph{Theorem 4.4 } Suppose we have two sets of real positive
numbers $$\{k_j\}, \quad \{B_j\}, \quad j=1,2,...,N, \quad k_j,
B_j>0,  \quad k_j, B_j \in \Real $$ such that numbers $k_j$ are
ordered in following way
$$k_1>k_2>...>k_N>0
$$
and $B_j$ are arbitrary. Let functions $\psi_1(x,k), \psi_2(x,k)$
have form
$$\psi_1(x,k)=e^{kx}(k^N+a_1k^{N-1}+...+a_N), \quad \psi_2(x,k)=(-1)^N\psi_1(x,-k)
$$
with indefinite real coefficients $a_j, j=1,2,...,N$.

Then there exist unique set of numbers $\{a_j\}$ such that two
functions $\psi_1(x,k_j), \ \psi_2(x,k_j)$ satisfy system of
equations
 \be \label{interp}
\psi_2(x,k_j)=(-1)^{j+1}B_j\psi_1(x,k_j)\ee and Wronskian of these
two functions has form \be
\label{wronadler}\mathcal{W}(k)=-2k\prod_1^N
(k^2-k_j^2).\ee\\

$\blacktriangleright$ Our first step is to construct $a_j$. It is
easy to see that (\ref{interp}) is equivalent to the following
system of equations on $\{a_j\}$ (for simplicity the system is
written out for a case $N=3$)

\be \label{adler}
\begin{cases}
k_1^2a_1+E_1k_1a_2+a_3+k_1^3E_1=0\\
k_2^2E_2a_1+k_2a_2+E_2a_3+k_2^3=0\\
k_3^2a_1+E_3k_3a_2+a_3+k_3^3E_3=0
\end{cases} \ee

where following notations has been used:
$$E_j=\frac{e^{\tau_j} -e^{-\tau_j}}{e^{\tau_j} +e^{-\tau_j}}= \tanh \tau_j,
\quad \tau_j=k_jx+\b_j, \quad B_j=e^{2\b_j}.$$ (To show this, it
is enough to write out explicitly $\psi_1$ and $\psi_2$ in roots
of polynomial and regard two cases: $N$ is odd and $N$ is even.
For instance, if $k=k_1$ and $N$ is odd, we get
$$
\psi_1(x,k)=e^{kx}(k^N+a_1k^{N-1}+...+a_N)= ...
\psi_2(x,k)=(-1)^N\psi_1(x,-k) ...
$$
)\\

Obviously, $0 \le E_j < 1$ and for a case $E_j=1, \forall j=1,2,3$
the system (\ref{adler}) takes form

\be \label{adler1}
\begin{cases}
k_1^2a_1+k_1a_2+a_3+k_1^3=0\\
k_2^2a_1+k_2a_2+a_3+k_2^3=0\\
k_3^2a_1+k_3a_2+a_3+k_3^3=0
\end{cases} \ee

(...). Thus, it was shown that determinant of Sys.(\ref{adler}) is
non-zero, i.e. all $a_j$ are uniquely defined and
functions $\psi_1(x,k_j), \ \psi_2(x,k_j)$ are polynomials.\\

In order to compute the Wronskian $\mathcal{W}$ of these two
functions, notice first that $\mathcal{W}$ is a polynomial with
leading  term $-2k^{2N+1}$.  Condition of proportionality (\ref{interp}) for functions
$\psi_1(x,k_j), \ \psi_2(x,k_j)$ provides that $k_j$ are zeros of the Wronskian and
that $\mathcal{W}$ is an odd function on $k$, i.e. (\ref{wronadler}) is proven. \qed\\

In order to illustrate  how to use this theorem for construction
of exact solutions with transparent potentials let us address  two
cases:  $N=1$ and $N=2$.

\paragraph{Example 4.5 } In case $N=1$ explicit form of functions
$$\psi_1=e^{kx}(k+a_1), \quad \psi_2=e^{-kx}(k-a_1)$$
allows us to find $a_1$ immediately:
$$a_1=-k_1E_1=-k_1 \tanh y_1= -k_1\tanh{(k_1x+\b_1)}$$
which coincides with formula for a solution of the same equation
obtained in Example 4.2
$$
\zeta_1 = -k_1\tanh{k_1(x-x_0)}
$$
for $x_0=\b_1/k_1.$ As to potential  $u$, it can be computed as
above using recurrent relation  which keeps true for all
$N$.\\

Notice that using this approach we have found solution of
Schrödinger equation by {\bf pure algebraic means} while in
Example 4.2  we had to solve Riccati equation in order to compute
coefficients of the corresponding truncated
series.  \\

The system (\ref{adler}) for case $N=2$ takes form

$$
\begin{cases}
k_1a_1+E_1a_2+k_1^2E_1=0\\
E_2k_2a_1+a_2+k_2^2=0
\end{cases} $$
which yields to

\bea \label{2soliton}
&& a_1=\frac{(k_2^2-k_1^2)E_1}{k_1-k_2E_1E_2}\nonumber \\
&& \ \ =D_x\log\left((k_2-k_1)\cosh(\tau_1+\tau_2)+
(k_2+k_1)\cosh(\tau_1-\tau_2)\label{2soliton} \right) \eea and
corresponding potential  $u=2a_{1,x}$ has explicit form
 \be\label{u2} u=
2 D_x^2\log\left((k_2-k_1)\cosh(\tau_1+\tau_2)+
(k_2+k_1)\cosh(\tau_1-\tau_2) \right) \ee where
$$x \to \pm \infty \quad \RA a_1 \to \pm
(k_1+k_2),$$ i.e. $u$ is a smooth function such that
$$u \to 0 \quad \mbox{for} \quad x \to \pm \infty. $$\\

Formulae (\ref{2soliton}) and (\ref{u2}) have been generalized for
the case of arbitrary $N$ by Hirota  whose
 work gave a
rise to a huge amount of papers dealing with construction of
soliton-like solutions for  {\bf nonlinear differential equations}
because some simple trick allows to add new variables in these
formulae (see \cite{kodama} and bibliography herein). For
instance, if we take
$$
\tau_j=k_jx+\b_j=k_jx+k_j^2y+k_j^3t
$$
then formula (\ref{u2}) gives particular solutions of
Kadomtzev-Petviashvili (KP) equation

\be \label{KP} (-4u_t+u_{xxx}+6uu_{x})_{x}+3u_{yy}=0 \ee

which is important model equation in the theory of surface waves.

\subsection{Finite-gap potentials}

In previous section it was shown how to construct integrable cases
of Schrödinger equation with soliton-like potentials vanishing at
infinity. Obvious - but not at all a trivial - next step is to
generalize these results for construction of integrable cases for
Schrödinger equation with {\bf periodic potentials}. In the
pioneering work \cite{Novikov1} the finite-gap potentials were
introduced and described in terms of their spectral properties but
deep discussion of spectral theory lays beyond  the scope of this
book (for exhaustive review see, for instance, \cite{Dubr1}). The
bottleneck of present theory of  finite-gap potentials is
following: spectral properties  formulated by Novikov´s school
provide only
 almost periodic potentials but do not guarantee periodic ones in
all the cases.\\

We are going to present here some simple introductory results
about finite-gap potentials and discuss a couple of examples. For
this purpose, most of the technique demonstrated in the previous
section can be used though as an auxiliary equation we will use
not Riccati equation but its
equivalent form, modified Schwarzian (\ref{modSchwarz}).\\

Generalization of Lemma 4.1 for the case of arbitrary finite
polynomial $c(x)$ can be formulated as follows.\\

\paragraph {Lemma 4.6} Equation for modified Schwarzian

 \be \label{modSchwarz1}
  \frac 34 \frac{h_x^2}{h^2}-\frac 12 \frac{h_{xx}}{h}+
\la^m h^2=U(x,\la):=\la^m+u_1 \la^{m-1}+...+ u_m
 \ee
with any polynomial generalized potential $U(x,\la)$ has unique
asymptotic solution represented by formal Laurant series such
that: \be \label{series1}
h(x,\la)=1+\sum_{k=1}^{\infty}\la^{-k}h_k(x) \ee where
coefficients $h_j$ are differential polynomials in all
$u_1,...,u_m$.\\

$\blacktriangleright$ The proof can be carried
out directly along the same lines as for Lemma 4.1. \qed\\

Direct corollary of this lemma is following: coefficients of the
formal solution are explicit functions of potential $U(x,\la)$. In
particular, for $m=1$ which corresponds to Schrödinger equation
(\ref{sch}) with generalized polynomial potential $\la + u$ we
have

\be \label{h1} h_1=\frac 12 u, \quad 2h_2= \frac 12
h_{1,xx}-h_1^2, ... \ee

\paragraph {Definition 4.7} Generalized potential
$$U(x,\la)=\la^m+u_1 \la^{m-1}+...+ u_m $$ of an equation
$$
\psi_{xx}=U(x,\la)\psi
$$
is called  {\bf $N$-phase potential} if Eq. (\ref{***}),
$$ \varphi_{xxx}=4U(x,\la)\varphi_{x}+2U_x(x)\varphi,$$
 has a
solution $\varphi$ which is a polynomial in $\la$ of degree $N$:

\be \label{gamma} \varphi(x,\la)=\la^N+\varphi_1(x) \la^{N-1}+...+
\varphi_N(x)=\prod_{j=1}^N (\la-\g_j(x)).\ee

Roots $\g_j(x)$ of the solution $\varphi(x,\la)$ are called {\bf
root variables}.\\

In particular case of Schrödinger equation this potential is also
called {\bf finite-gap potential}. As it follows from
\cite{Dubr1}, original "spectral" definition of a finite-gap
potential is equivalent to our Def. 4.7 which is more convenient
due to its applicability not only for Schrödinger equation but
also for arbitrary equation of the second order.\\

There exists direct connection between the statement of Lemma 4.6
and the notion of finite-gap potential. One can check directly
that if function $h(x,\la)$ is solution of Eq.(\ref{modSchwarz1}),
then function $\varphi(x,\la) = 1/h(x,\la)$ is solution of
Eq.(\ref{***}) and can be written as formal series

\be \label{series2}
\varphi(x,\la)=1+\sum_{k=1}^{\infty}\la^{-k}\varphi_k(x) \ee

with coefficients which are explicit functions of generalized
potential $U(x,\la)$. In case when series (\ref{series2}) becomes
a finite sum, we get finite-gap potential.\\

\paragraph {Example 4.8} Let a solution $\varphi(x, \la)=\la-\g (x)$ is a
polynomial of first degree and potential is also linear, i.e.
$m=1, \ N=1$. Then after integrating the equation from definition
above, we get \be \label{**1}
  4(\la + u)\varphi^2+\varphi_x^2-2\varphi \varphi_{xx}= c(\la)
\ee with some constant of integration $c(\la)$ and left part of
(\ref{**1}) is a polynomial in $\la$ of degree 3,

\be
\label{polC}C(\la)=4\la^3+c_1\la^2+c_2\la+c_3=4(\la-\la_1)(\la-\la_2)(\la-\la_3),
 \ee
where $\la_i$ are all roots of the polynomial $C(\la)$.
Eq.(\ref{**1}) is identity on $\la$ and therefore without loss of
generality we write further $C(\la)$ for both sides of it. This
identity has to keep true for all values of $\la$, in particular,
also  for $\la=\g (x)$ which gives \be
\label{ellipt}\g^2_x=C(\g)=4(\g-\la_1)(\g-\la_2)(\g-\la_3).\ee

Now instead of solving Eq.(\ref{**1}), we have to solve
Eq.(\ref{ellipt}) which is
 integrable in elliptic functions.\\

If we are interested in real solutions without singularities, we
have to think about initial data for Eq.(\ref{ellipt}). For
instance, supposing that all $\la_j$ are real,  without loss of
generality
$$ \la_1 > \la_2 > \la_3, $$ and for  initial data $(x_0,\g_0)$ satisfying
 $$ \forall (x_0,\g_0) : \ \la_3<\g_0<\la_2,$$
Eq.(\ref{ellipt}) has real smooth periodic solution expressed in
elliptic functions \be
\label{elliptPoten}u=2\g-\la_1-\la_2-\la_3\ee
 It is our finite-gap potential
(1-phase potential) and its period can be computed explicitly as
$$
T=\int_{\la_3}^{\la_2}\frac{d\la}{\sqrt{(\la-\la_1)(\la-\la_2)(\la-\la_3)}}.
$$

We have regarded in Example 4.8 particular case $mN=1$. Notice
that in general case of $mN>1$  following the same reasoning,
after integration we get  polynomial $C(\la)$ of degree $2N+m$ \be
\label{**2}
  4U(x,\la)\varphi^2+\varphi_x^2-2\varphi \varphi_{xx}=
C(\la):=4\la^{2N+m}+... \ee and correspondingly a system of
$2N+m-1$ equations on $N$ functions, i.e. the system will be
over-determined.  On the other hand, choice of $\la=\g_j$ makes it
possible to get a closed subsystem of $N$ equations for $N$
functions as above: \be \label{gammmN} \g^2_{j,x}=C(\g_j)/\prod_{j
\neq k} (\g_j-\g_k)^2. \ee Following Lemma shows that this
over-determined system of equations has unique solution which is
defined by Sys.(\ref{gammmN}).

\paragraph {Dubrovin´s Lemma} Let system of differential equations
(\ref{gammmN}) on root variables $\g_j$ defined by (\ref{gamma})
is given with
$$
C(\la) = 4\la^{2N+m}+...,
$$
then following keeps true:\\
1. $C(\la)|_{\la=\g_j}=\varphi_x^2(x,\la)|_{\la=\g_j}, \
j=1,...,N$,\\
2. expression
$$\varphi^{-1}(2\varphi_{xx}+\frac{C(\la)-\varphi^2_x}{\varphi})$$
is a polynomial in $\la$ of degree $m$ and leading coefficient
1.\\

$\blacktriangleleft$ Let us notice that
$$\big( \prod(\la - \g_k) \big)_x |_{\la=\g_j}=\big( -\g_{1,x}\prod_{j=2}^N(\la-\g_j)-....-
\g_{N,x}\prod_{j=1}^{N-1}(\la-\g_j)\big)|_{\la=\g_j}$$$$=-\g_{j,x}\prod_{j\neq
k }(\g_j-\g_k)
$$ which implies
$$
\varphi_x|_{\la=\g}= -\g_{j,x} \prod_{j\neq k }(\g_j-\g_k) \RA
C(\la)|_{\la=\g_j}=\varphi_x^2(x,\la)|_{\la=\g_j},
$$
i.e. first statement of the lemma is proven.\\

(...). \qed


 Below it will be shown that also {\bf
transparent potentials themselves} can be computed algebraically.
(...)


\section{Summary}

In this Chapter, using Riccati equation as  our main example,
 we tried to demonstrate at least some of
the ideas and notions introduced in Chapter 1 - integrability in
quadratures, conservation laws, etc. Regarding transformation
group and singularities of solutions for RE,  we  constructed some
equivalent forms of Riccati equation. We also compared three
different approaches to the solutions of Riccati equation and its
equivalent forms. The classical form of RE allowed us to construct
easily asymptotic solutions represented by formal series. Linear
equation of the second order turned out to be more convenient to
describe finite-gap potentials for exact solitonic solutions which
would be a much more complicated task for a RE itself while
generalization of soliton-like potentials to finite-gap potentials
demanded modified Schwarzian equation. \\

In our next Chapter we will show that modified Schwarzian equation
also plays important role in the construction of a differential
operator commuting with a given one while existence of commuting
operators allows us to obtain examples of hierarchies for
solitonic equations using Lemma 4.6. In particular, for $m=1$
coefficients $h_k(x)$ of Eq.(\ref{series1}) describe a set of
conservation laws for {\bf Korteweg-de Vries equation} ({\bf KdV})

\be \label{KdV} u_t+6uu_x+u_{xxx}=0. \ee

\section{Exercises for Chapter 2}
\paragraph{1.} Prove that general solution of $z^{'}=a(x)z$ has a
form $$z(x)=e^{\int a(x)dx}.$$

\paragraph{2.} Deduce formula (\ref{psi2}) regarding $<\hat{\psi}_1,
\hat{\psi}_2>=1$ as a linear first order equation on
$\hat{\psi}_2$.


\paragraph{3.} Prove that for $L=\frac{d^m}{dx^m}$ its kernel
consists of all polynomials of degree $\leq m-1$.

\paragraph{4.} Let functions $A_1$ and $A_2$ are two solutions of
(\ref{***}).  Prove that the Wronskian $<A_1,\,
A_2>=A_1A_2'-A_2A_1'$ is solution as well.

\paragraph{5.} Let the function $A$ satisfies (\ref{**}). Prove that the
functions
\[   f_\pm=\frac12D\log A\pm\frac{\sqrt z}{A}  \]
satisfies the Riccati equations (\ref{ric}).

\paragraph{6.} Proof that
$$\frac 34 \frac{a_x^2}{a^2}-\frac 12 \frac{a_{xx}}{a}=k^2
\LRA a=(\eps_1e^{kx}+\eps_2e^{-kx})^{-2}. $$

}

\end{document}